\documentclass[prl, aps, twocolumn, superscriptaddress]{revtex4}
\usepackage{amssymb,amsthm,amsmath}

\theoremstyle{plain}
\newtheorem{theorem}{Theorem}

\newtheorem{corollary}{Corollary}
\newtheorem{proposition}{Proposition}

\theoremstyle{definition}

\newcommand{\be}{\begin{equation}}
\newcommand{\ee}{\end{equation}}
\newcommand{\bea}{\begin{eqnarray}}
\newcommand{\eea}{\end{eqnarray}}
\newcommand{\Tr}{\text{tr}\;}

\newcommand{\spec}{\text{spec}}

\newcommand{\cU}{{\cal U}}      
\newcommand{\cV}{{\cal V}}      
\newcommand{\cS}{{\cal S}}      
\newcommand{\cB}{{\cal B}}      
\newcommand{\cC}{\mathbb{C}}    
\newcommand{\cR}{{\cal R}}
\begin{document}

\title{The Spectra of Quantum States and the \\Kronecker Coefficients of the Symmetric Group}

\author{Matthias \surname{Christandl}}
\email[]{matthias.christandl@qubit.org}
\affiliation{Centre for
Quantum Computation, Department of Applied Mathematics and
Theoretical Physics, University of Cambridge, Wilberforce Road,
Cambridge, CB3 0WA, United Kingdom}
\author{Graeme \surname{Mitchison}}
\email[]{G.J.Mitchison@damtp.cam.ac.uk}
\affiliation{Centre for
Quantum Computation, Department of Applied Mathematics and
Theoretical Physics, University of Cambridge, Wilberforce Road,
Cambridge, CB3 0WA, United Kingdom}
\affiliation{MRC Laboratory of
Molecular Biology, University of Cambridge, Hills Road, Cambridge,
CB2 2QH, United Kingdom}

\date{\today}

\begin{abstract}
Determining the relationship between composite systems and their
subsystems is a fundamental problem in quantum physics. In this
paper we consider the spectra of a bipartite quantum state and its
two marginal states. To each spectrum we can associate a
representation of the symmetric group defined by a Young diagram
whose normalised row lengths approximate the spectrum. We show
that, for allowed spectra, the representation of the composite
system is contained in the tensor product of the representations
of the two subsystems. This gives a new physical meaning to
representations of the symmetric group. It also introduces a new
way of using the machinery of group theory in quantum
informational problems, which we illustrate by two simple
examples.
\end{abstract}

\maketitle
\subsection*{I. INTRODUCTION}

In 1930, Weyl observed with dry humour that the ``group
pest'' seemed to be here to stay (\cite{Weyl50}, preface to second
German edition). The theory of representations of groups, which he
did so much to develop, is indeed a firmly established component
of modern physics, appearing wherever the relation of a composite
system to its parts is investigated. The aim of this paper is to
derive a novel connection between certain representations and the
properties of composite quantum systems.

Suppose a quantum system consists of two parts, $A$ and $B$, and
let $\rho^{AB}$ be a density operator on the composite system
$AB$. The states $\rho^A$ and $\rho^B$ obtained by tracing out the
subsystems $B$ and $A$, respectively, are constrained by the fact
that they are derived from a common state. For instance,
subadditivity and the triangle inequality are informational
inequalities that relate the von Neumann entropies (the Shannon
entropies of the spectra) of $\rho^{AB}$, $\rho^A$ and $\rho^B$.
Even more fundamentally, however, one can ask what constraints
there are on the spectra of $\rho^A$ and $\rho^B$ once one knows
the spectrum of $\rho^{AB}$. We prove here a theorem that relates
this problem to certain representations of the unitary and
symmetric groups.

A familiar example of a composite system is two particles, one
with spin $j_1$ and the other with spin $j_2$. The addition of
their angular momenta can be described in terms of representations
of SU(2); the product of two representations, one for each
subsystem, can be expressed as a sum of representations on the
total system. This is the familiar Clebsch-Gordan series, whose
coefficients have been much studied and can be readily calculated.
There is an analogous expansion of the product of two
representations of the symmetric group $S_k$ on $k$ elements. The
coefficients appearing in this alternative Clebsch-Gordan series
are known as Kronecker coefficients, and their evaluation is more
difficult: no simple algorithm is known at present.

To state our result, we need a little notation. As we shall see
shortly, every irreducible representation of the symmetric group
$S_k$ can be labelled by an ordered partition $\lambda=(\lambda_1,
\ldots, \lambda_q)$ of $k$; i.e. a set of non-negative integers
$\lambda_i$ with $\lambda_{i+1} \leq \lambda_i$ and $\sum
\lambda_i=k$. Let $\bar \lambda$ denote $(\frac{\lambda_1}{k},
\ldots, \frac{\lambda_q}{k})$ and let $g_{\lambda \mu\nu}$ denote
the Kronecker coefficient that counts the number of times
(possibly zero) that the representation labelled by $\lambda$
appears in the product of those labelled by $\mu$ and $\nu$.
Finally, let $\spec (\rho)$ denote the spectrum of $\rho$. Then
our main result is that, given a density operator $\rho^{AB}$ with
$\spec (\rho^{AB})=\bar \lambda$, $\spec (\rho^A)=\bar \mu$ and
$\spec (\rho^B)=\bar \nu$, there is a sequence $\lambda_j, \mu_j,
\nu_j$ with non-zero $g_{\lambda_j \mu_j \nu_j}$ such that
$\bar{\lambda}_j, \bar{\mu}_j$ and $\bar{\nu}_j$ converge to
$\spec (\rho^{AB}), \spec (\rho^A)$ and $\spec (\rho^B)$,
respectively.

\subsection*{II. YOUNG DIAGRAMS AND THE SPECTRUM OF A DENSITY OPERATOR}

In this section we give a brief description of representations of the
symmetric group $S_k$ and the special unitary group in $d$ dimensions,
$SU(d)$, and review a theorem by Keyl and Werner \cite{KW01}, which
will play a key role in proving our main result.

If ${\cC}^d$ denotes a $d$-dimensional complex vector space, $S_k$
operates on $({\cC}^d)^{\otimes k}$ by
\be \label{equation-symmetry-action}\pi \{e_{i_1}
\otimes e_{i_2} \otimes \ldots \otimes e_{i_k}\}=
e_{i_{\pi^{-1}(1)}} \otimes e_{i_{\pi^{-1}(2)}} \otimes \ldots
\otimes e_{i_{\pi^{-1}(k)}},
\ee for $\pi
\in S_k$, where the $e_1, \ldots e_d$ are elements of some basis
of ${\cC}^d$. The group $SU(d)$ acts by
\be \label{equation-SU-action} e_{i_1} \otimes e_{i_2}
\otimes \ldots \otimes e_{i_k} \to Ue_{i_1} \otimes Ue_{i_2}
\otimes \ldots \otimes Ue_{i_k},\ee for $U \in SU(d)$.

 These actions of $S_k$ and
$SU(d)$ on $({\cC}^d)^{\otimes k}$ define representations of each
group, but both representations are reducible. Their irreducible
components can be constructed as follows. Let us write $\lambda
\vdash k$ to mean that $\lambda$ is an ordered partition with
$\sum \lambda_i=|\lambda|=k$. This can be depicted by a
\emph{Young frame}, which consists of $q$ rows, the $i$-th row
having $\lambda_i$ boxes in it. A \emph{Young tableau} $T$ is
obtained from a frame by filling the boxes with the numbers $1$ to
$k$ in some order, with the constraint that the numbers in each
row increase on going to the right and the numbers in each column
increase downwards.

To each tableau $T$, we associate the \emph{Young symmetry
operator} $e(T)$ given by
\be \label{symmetry-operator} e(T)=\Big(\sum_{\pi \in {\cal C}(T)}
\text{sgn } (\pi) \pi \Big)\Big(\sum_{\pi \in {\cal R}(T)}
\pi\Big),
\ee where ${\cal R}(T)$ and ${\cal C}(T)$ are sets of permutations
of $S_k$, ${\cal R}(T)$ being those that are obtained by permuting
the integers within each row of $T$, and ${\cal C}(T)$ those
obtained by permuting integers within each column of $T$
\cite{Weyl50}.

Each $e(T)$ satisfies $e(T)^2=re(T)$ for some integer $r$, so
$e(T)/r$ is a projection which we denote by $p(T)$. The action of
$SU(d)$ on the image subspace of $p(T)$ in $({\cC}^d)^{\otimes k}$
gives an irreducible representation of $SU(d)$. If $T'$ is another
tableau of the same frame, the representations of $SU(d)$ are
equivalent (under the permutation that takes $T$ to $T'$). Thus
the irreducible representations of $SU(d)$ are labelled by Young
frames, or equivalently, by partitions $\lambda \vdash k$.

Now pick a vector $v$ in the subspace defined by $p(T)$, and apply
all elements $\pi \in S_k$ to it. The subspace of
$({\cC}^d)^{\otimes k}$ spanned by $\{\pi v: \pi \in S_k\}$
defines an irreducible representation of $S_k$. Distinct
frames yield distinct representations, so we can also label the
irreducible representations of $S_k$ by partitions $\lambda \vdash
k$.

From the above construction, it can be shown that the subspaces
$\cU_\lambda$ and $\cV_\lambda$ of the irreducible representations
of $S_k$ and $SU(d)$, respectively, are related in the following
elegant manner:
\be \label{equation-duality} \left( {\cC}^d \right)^{\otimes
k}=\bigoplus_{\lambda \vdash k} \cU_\lambda \otimes \cV_\lambda.
\ee This is sometimes called the Weyl-Schur duality of $S_k$ and
$SU(d)$.

A systematic way to generate ${\cal V}_\lambda$ for a tableau $T$
is to apply $p(T)$ to all vectors $v=e_{i_1} \otimes e_{i_2}
\otimes \ldots \otimes e_{i_k}$, where we identify the $j$-th
component of the tensor product with the $j$-th box in the
numbering of the tableau $T$. If we count the number of times each
basis element $e_i$ occurs in $v$, this defines a partition $\nu
\vdash k$. We say $\nu$ is {\em majorized} by $\lambda$, and write
$\nu \prec \lambda$, if $\sum_{i=1}^q \nu_i \le \sum_{i=1}^q
\lambda_i$ for $q=1, \ldots , d-1$ and $\sum_{i=1}^d \nu_i =
\sum_{i=1}^d \lambda_i$. The vector $v$ will project to zero under
$p(T)$ unless $\nu \prec \lambda$, since otherwise there must be
two boxes in the same column of $T$, with numberings $i$ and $j$,
for which $e_i=e_j$. In particular, for any Young diagram with
more than $d$ rows, ${\cal V}_\lambda=0$.

The dimensions of $\cV_\lambda$ and $\cU_\lambda$ are given by
\cite{IN66}
\be \dim {\cal V}_\lambda = \frac{\prod_{i<j}
(\lambda_i-\lambda_j-i+j)}{\prod_{m=1}^{d-1} m!}.  \ee and
\be \dim {\cal U}_\lambda= \frac{k!}{\prod_i \mbox{hook}(i)}, \ee
where the index $i$ in the latter formula runs over all boxes in
the Young diagram of $\lambda$, and $\mbox{hook}(i)$ is the
hook-length of box $i$, i.e. the number of boxes vertically below
$i$ and to the right of $i$ within the diagram, including box $i$.
Useful bounds for these dimensions are
\be \label{Vbound} \dim {\cal V}_\lambda \le (k+1)^{d(d-1)/2} \ee
and
\be \label{Ubound} \frac{k!}{\prod_i (\lambda_i+d-i)!} \le \dim
{\cal U}_\lambda \le \frac{k!}{\prod_i \lambda_i!}. \ee

A remarkable connection between Young frames and density operators
was discovered by Keyl and Werner \cite{KW01} (see also R. Alicki,
S. Rudnicki and S. Sadowski~\cite{AlRuSa87}). Suppose $\rho$ is a
density operator whose spectrum $\spec(\rho)$. Keyl and Werner
showed that, for large $k$, the quantum state $\rho^{\otimes k}$
will project with high probability into the Young subspaces $\lambda
\vdash k$ such that $\bar{\lambda}$ approximates $\spec(\rho)$.
Their proof was somewhat elaborate. A succinct argument was found by
Hayashi and Matsumoto \cite{HM02}, and we give it here, correcting
an algebraic slip in their derivation.

\begin{theorem} \label{theorem-Keyl-Werner}
Let $\rho$ be a density operator with spectrum $r=\spec(\rho)$, and
let $P_\lambda$ be the projection onto $\cU_\lambda \otimes
\cV_\lambda$. Then \be \label{eq-KeylWerner-1}\Tr P_\lambda
\rho^{\otimes k} \leq (k+1)^{d(d-1)/2} \exp \left(-k D
(\bar{\lambda}||r)\right)\ee with $D(p||q)=\sum_i p_i (\log p_i
-\log q_i)$ the \emph{Kullback-Leibler distance} of two normalised
probability distributions $p$ and $q$. Note that $D(p||q)=0$ if and
only if $p=q$.
\end{theorem}

\begin{proof}
In the procedure for generating ${\cal V}_\lambda$ described
above, we can choose the eigenvectors of $\rho$ as a basis for
${\cC}^d$. When eigenvectors of $\rho^{\otimes k}$ are projected
onto $\cU_\lambda \otimes \cV_\lambda$, the result is non-zero
only if $\rho \prec\lambda$, as argued above. The `surviving'
eigenvalues $\prod_i r_i^{\mu_i}$, where $\mu \vdash k$, are
therefore smaller than $\prod_i r_i^{\lambda_i}$.

Using the bounds (\ref{Vbound}) and (\ref{Ubound}), it follows
that
\bea \Tr P_\lambda \rho^{\otimes k} &\leq& \dim {\cal U}_\lambda \dim
{\cal V}_\lambda \prod_i r_i^{\lambda_i} \\
&\leq& (k+1)^{d(d-1)/2} \frac{k!}{\prod_i \lambda_i!}\prod_i
r_i^{\lambda_i} \\ &\leq& (k+1)^{d(d-1)/2} \exp \left(-k D
(\bar{\lambda}||r)\right). \eea This completes the proof.
\end{proof}

\begin{corollary}
\label{eq-KeylWerner2} If $\rho$ is a density operator with spectrum
$r=\spec(\rho)$, \be \Tr P_X \rho^{\otimes k} \leq (k+1)^{d(d+1)/2}
\exp (-k \min_{\lambda \vdash n: \bar{\lambda} \in {\cal S}}D
(\bar{\lambda}||r)), \ee where $P_X:= \sum_{\lambda \vdash k:
\bar{\lambda} \in {\cal S}}P_\lambda$ for a set of spectra $\cS$ .
\end{corollary}

This follows from the theorem if we simply pick the Young frame
with the slowest convergence and multiply it by the total number
of possible Young frames with $k$ boxes in $d$ rows. This number
is
certainly smaller than $(k+1)^d$. \\

Let ${\cal B}_\epsilon(r):=\{r': \sum |r'_i-r_i|< \epsilon\}$ be
the $\epsilon$-ball around the spectrum $r$. If we take ${\cS}$ to
be the complement of ${\cB}_\epsilon(r)$, it becomes clear that
for large $k$, $\rho^{\otimes k}$ will project into a Young
subspace $\lambda$ with $\bar{\lambda}$ close to $r$ with high
probability. More precisely

\begin{corollary}
\label{corollary-Keyl-2} Given an operator $\rho$ with spectrum
$r=\spec(\rho)$, and given $\epsilon_1>0$, let $P_X=\sum_{\lambda
\vdash k: \bar{\lambda} \in \cB_{\epsilon_1}(r)} P_\lambda$. Then
for any $\epsilon_2>0$ there is a $k_0>0$ such that for all $k\geq
k_0$, \be \Tr P_X \rho^{\otimes k}>1-\epsilon_2. \ee
\end{corollary}

\subsection*{III. THE CONTENT EXPANSION}

Suppose now we have a bipartite system $AB$ made up of systems $A$
and $B$ with spaces $\cC^m$ and $\cC^n$, respectively. $SU(mn)$ thus
acts on $\cC^{mn}$, the space of $AB$, and hence on the $k$-fold
tensor product $(\cC^{mn})^{\otimes k}$ according to equation
(\ref{equation-SU-action}). Similarly, $SU(m)$ and $SU(n)$ act on
$A$ and $B$, respectively, this gives an action of $SU(m) \times
SU(n)$ on $AB$. If $\cR_\lambda$ is an irreducible representation of
$SU(mn)$ on the Young subspace $\lambda$ of $AB$, its restriction to
$SU(m) \times SU(n)$ is not necessarily irreducible, and can
generally be expressed as a sum of terms $\cR_\mu \otimes \cR_\nu$,
where $\cR_\mu$ and $\cR_\nu$ are irreducible representions of
$SU(m)$ and $SU(n)$, respectively. We call this sum the {\em content
expansion} of $\lambda$, borrowing some terminology from
\cite{HM65:1,IN66}.

In the remainder of this section we follow \cite{Macdonald79}
closely. The content expansion can be conveniently described in
terms of the characters of the underlying representations. The
conjugacy class of a unitary matrix in $SU(d)$, for some dimension
$d$, is given by its $d$ eigenvalues $x_1, \ldots ,x_d$. The
character $s_\lambda$ of the representation $\cV_\lambda$ of
$SU(d)$ is therefore a function of $x_1, \ldots ,x_d$. Define the
homogeneous power sums by $h_r=\sum_{i_1 \leq i_2 \leq \cdots \leq
i_r} x_{i_1}x_{i_2} \cdots x_{i_r}$. Then \be s_\lambda(x)=\det
(h_{\lambda_i-i+j})_{1 \leq i,j \leq d}.  \ee The polynomial
$s_\lambda(x)$ is called the \emph{Schur function} or
\emph{S-function} of $\lambda$. The $s_\lambda$ form a basis for
the symmetric polynomials in $d$ variables of degree $|\lambda|$.

Now take the character $s_\lambda(z)=s_\lambda(z_1, \ldots ,
z_{mn})$ of the representation $\lambda$ of $SU(mn)$. When
restricted to $SU(m) \times SU(n)$, this can be regarded as the
function $s_\lambda(xy)$, where $x_1, \ldots x_m$ are eigenvalues of
an element of $SU(m)$, and $y_1, \ldots y_n$ those of an element of
$SU(n)$, and $xy$ denotes the set of all products $x_iy_j$. The
products $s_\mu(x)s_\nu(y)$ of Schur functions over all $\mu$ and
$\nu$ with $|\mu|=|\nu|=|\lambda|$ are the characters of the
irreducible representations of $SU(m) \times SU(n)$. Hence we can
write $s_\lambda(xy)$ in this basis, and obtain the content
expansion: \be \label{product-expansion}
s_\lambda(xy)=\sum_{\mu,\nu} g_{\lambda\mu\nu}s_\mu(x)s_\nu(y). \ee

The relationship between representations of $SU(d)$ corresponding to
eq. (\ref{product-expansion}) can be written \be
\label{equation-fine-graining} {\cR}_\lambda \downarrow_{SU(m)
\times SU(n)}=\bigoplus_{\mu, \nu} g_{\lambda \mu \nu} {\cR}_\mu
\otimes {\cR}_\nu, \ee where the left hand side denotes the
representation $\cR_\lambda$ of $SU(mn)$ restricted to the subgroup
$SU(m)\times SU(n)$. Note that the underlying subspace of $\cR_\mu
\otimes \cR_\nu$ is not in general $\cV_\mu \otimes \cV_\nu$, but is
embedded in $(\cU_\mu \otimes \cV_\mu) \otimes (\cU_\nu \otimes
\cV_\nu)$ by some unitary action.

The integers $g_{\lambda\mu\nu}$ are sometimes called the {\em Kronecker
coefficients}; this name alludes to another context they occur in, as we now
explain.

Let $\chi_\lambda (\tau)$ denote the character of the
representation $\cU_\lambda$ of $S_k$ on the conjugacy class
$\tau$. The conjugacy class of a permutation in $S_k$ is
determined by the lengths of the cycles in the permutation, so
$\tau$ is a partition of $k$ whose parts $\tau_i$ represent the
lengths of those cycles. Let $p_r(x_1, \dots , x_d)=\sum_i x_i^r$
be the $r$th \emph{power sum} and let $p_\lambda=p_{\lambda_1}
p_{\lambda_2} \cdots p_{\lambda_q}$. Then the character of the
representation of $SU(d) \times S_k$ given by equations
(\ref{equation-symmetry-action}) and (\ref{equation-SU-action})
takes the value $p_\tau(x)=p_\tau(x_1, \dots , x_d)$ at an element
of $SU(d)$ with eigenvalues $x_1, \dots , x_d$ and a permutation
with class $\tau \vdash k$. In terms of characters, therefore,
eq.(\ref{equation-duality}) becomes \be
\label{equation-duality-sym} p_\tau(x)=\sum_\lambda \chi_\lambda
(\tau) s_\lambda(x).\ee Thus \be \label{product1}
p_\tau(x)p_\tau(y)=\sum_{\mu\nu} \chi_\mu (\tau)\chi_\nu (\tau)
s_\mu(x) s_\nu(y), \ee but since $p_\tau(x) p_\tau(y)=p_\tau(xy)$,
by equation (\ref{product-expansion}) we have
\be \label{product2} p_\tau(x)p_\tau(y)=\sum_{\lambda \mu \nu}
g_{\lambda\mu\nu} \chi_\lambda(\tau) s_\mu(x)s_\nu(y).
\ee Comparing equations (\ref{product1}) and (\ref{product2}), and noting that
the products $s_\mu(x)s_\nu(y)$ are linearly independent, we have
for each $\mu$
\be \label{Symmetric-CG} \chi_\mu (\tau) \chi_\nu
(\tau) = \sum_{\lambda} g_{\lambda\mu\nu}\chi_\lambda(\tau),
\ee
with $|\lambda|=|\mu|=|\nu|$. This implies that the corresponding
representation subspaces satisfy
\be
\label{Symmetric-CG-subspaces} \cU_\mu \otimes \cU_\nu
=\bigoplus_{\lambda} g_{\lambda \mu \nu} \cU_\lambda.
\ee

Thus the Kronecker coefficients $g_{\lambda\mu\nu}$ appear in the
Clebsch-Gordan series for the symmetric group $S_k$, i.e. in the
series that represents the Kronecker (or tensor) product of two
irreducible representations of $S_k$ as a sum of irreducible
representations, where the latter are weighted by the number of
times they occur. This is analogous to the Clebsch-Gordan series for
$SU(d)$, \be s_\mu (x) s_\nu (x) = \sum_{\lambda \mu \nu}
c_{\mu\nu}^\lambda s_\lambda(x), \ee where $|\lambda|=|\mu|+|\nu|$.
The coefficients $c_{\mu\nu}^\lambda$ can be calculated using the
famous Littlewood-Richardson rule \cite{Macdonald79}, whereas
finding an efficient computational rule for the $g_{\lambda\mu\nu}$
is a fundamental open problem (see e.g. \cite{Kir04}).

Equation (\ref{Symmetric-CG}) and orthogonality of characters
implies
\be g_{\lambda \mu \nu}=\frac{1}{k!} \sum_{\tau \in S_k}
\chi_\lambda(\tau) \chi_\mu (\tau)\chi_\nu (\tau),
\label{symmetry}
\ee
which shows that the Kronecker coefficients are symmetric under
interchange of the indices.

\subsection*{IV. MAIN RESULT}

We now show that there is a close correspondence between
the spectra of a density operator $\rho^{AB}$ and its traces $\rho^A$ and
$\rho^B$ and Young frames $\lambda$, $\mu$, $\nu$ with positive Kronecker
coefficients. More precisely,

\begin{theorem}
For every density operator $\rho^{AB}$, there is a sequence
$(\lambda_j, \mu_j, \nu_j)$ of partitions, labelled by natural numbers
$j$, with $|\lambda_j|=|\mu_j|=|\nu_j|$ such that \be g_{\lambda_j
\mu_j \nu_j} \neq 0 \quad \mbox{ for all }\ j\ee and
\bea \lim_{j \to
\infty} \bar{\lambda}_j=\spec (\rho^{AB})\\ \lim_{j \to \infty}
\bar{\mu}_j=\spec (\rho^{A})\\ \lim_{j \to \infty} \bar{\nu}_j=\spec
(\rho^{B}) \eea
\label{ourtheorem}
\end{theorem}

\begin{proof}
Let $r^{AB}=\spec(\rho^{AB}), r^{A}=\spec(\rho^{A}),
r^{B}=\spec(\rho^{B})$. Let $P^{AB}_\lambda$ denote the projector
onto the Young subspace $\cU_\lambda \otimes \cV_\lambda$ in
system $AB$, and let $P^A_\mu$, $P^B_\nu$ be the corresponding
projectors onto Young subspaces in $A$ and $B$, respectively. By
corollary \ref{corollary-Keyl-2}, for given $\epsilon>0$, we can
find a $k_0$ such that the following all hold for all $k \geq
k_0$,
\bea \label{equation-sum1} \Tr P_X (\rho^A)^{\otimes k} &\geq& 1-\epsilon, \quad P_X:=\sum_{\bar{\mu} \in
\cB_\epsilon(r^A)} P^{A}_\mu\\
 \label{equation-sum2}\Tr P_Y (\rho^B)^{\otimes k} &\geq& 1-\epsilon, \quad P_Y:=\sum_{\bar{\nu} \in \cB_\epsilon(r^B)}
P^{B}_\nu\\
\label{equation-sum3} \Tr P_Z (\rho^{AB})^{\otimes k} &\geq&
1-\epsilon, \quad P_Z:=\sum_{\bar{\lambda} \in
\cB_\epsilon(r^{AB})} P^{AB}_\lambda.
\eea
Equations (\ref{equation-sum1}) and (\ref{equation-sum2}) can be
combined to yield
\be \label{equation-sum4} \Tr (P_X \otimes P_Y) (\rho^{AB})^{\otimes k} \geq
1-2\epsilon.
\ee
This follows from
\be \label{trace-inequality} \Tr (P \otimes Q) \xi^{AB}
\ge \Tr P \xi^A + \Tr Q \xi^B-1, \ee which holds for all projectors $P$ and
$Q$ and density operators $\xi^{AB}$ since $\Tr [(\openone -P) \otimes
(\openone-Q) \xi^{AB} ]\geq 0$.

Because $(\rho^{AB})^{\otimes k}$ maps each Young frame into itself, writing
$\sigma=(\rho^{AB})^{\otimes k}$, we have
\be \label{eq-project}\sum_{\lambda \vdash k} P_\lambda
\sigma P_\lambda=\sigma.
\ee
Defining $P_{\bar Z}:=\openone-P_Z$, eqs. (\ref{equation-sum4})
and (\ref{eq-project}) imply
\be \Tr [(P_X \otimes P_Y) (P_Z\sigma P_Z + P_{\bar Z}\sigma P_{\bar
Z})] \geq 1-2 \epsilon.
\ee
We now insert $\Tr [(P_X \otimes P_Y) P_{\bar Z}\sigma P_{\bar Z}] \leq
\epsilon$ (from eq. (\ref{equation-sum3})) and obtain
\be \Tr [(P_X \otimes P_Y) P_Z\sigma P_Z] \geq 1-3 \epsilon.
\ee

Clearly, there must be at least one triple $\mu \in \cB_\epsilon(r^A), \nu
\in \cB_\epsilon(r^B)$ and $\lambda \in \cB_\epsilon(r^{AB})$ with
$\Tr[(P^A_\mu \otimes P^B_\nu) P^{AB}_\lambda \sigma P^{AB}_\lambda] \neq 0$.
Thus $(P^A_\mu \otimes P^B_\nu) P^{AB}_\lambda \neq 0$ and by eq.
(\ref{equation-fine-graining}) this implies $g_{\lambda \mu \nu} \neq 0$.
\end{proof}

Next we consider some consequences of the above theorem. The von
Neumann entropy $S(\rho)$, of the operator $\rho$, is defined by
$S(\rho)=-\Tr(\rho \log(\rho))=H(r)$, where $r$ is the spectrum of
$\rho$.

\begin{proposition} Von Neumann entropy is subadditive; i.e. for all
$\rho^{AB}$, $S(\rho^{AB}) \le S(\rho^A) + S(\rho^B)$.\\
\end{proposition}

\begin{proof}
The Clebsch-Gordan expansion for the symmetric group, eq.
(\ref{Symmetric-CG-subspaces}), implies
\be
g_{\lambda \mu \nu} \dim {\cal U}_\lambda \le \dim {\cal U}_\mu
\dim {\cal U}_\nu.
\ee
Theorem \ref{ourtheorem} tells us that, for every operator
$\rho^{AB}$, there is a sequence of non-vanishing $g_{\lambda_j
\mu_j \nu_j}$ with $\bar \lambda_j$, $\bar \mu_j$, $\bar \nu_j$
converging to the spectra of $\rho^{AB}$, $\rho^A$ and $\rho^B$.
Since the Kronecker coefficients are always non-negative integers,
if $g_{\lambda_j \mu_j \nu_j} \ne 0$, we have
\be \label{space2}
\dim \cU_{\lambda_j} \le \dim \cU_{\mu_j} \dim \cU_{\nu_j}.
\ee
This holds for all $j$, and in the limit of large $j$ Stirling's
approximation and inequality (\ref{Ubound}) imply that $\log(\dim
\cU_{\lambda_j})$ tends to $k S(\rho^{AB})$, where
$k=|\lambda_j|$, and similarly for systems $A$ and $B$. Thus we
obtain subadditivity.

\end{proof}

\begin{proposition} The triangle inequality \cite{AL70},
$S(\rho^{AB}) \ge |S(\rho^A) - S(\rho^B)|$ holds for all $\rho^{AB}$.
\end{proposition}
\begin{proof} The symmetry of the coefficients implied by eq. (\ref{symmetry}) tells us that in addition to
eq. (\ref{space2}) we also have the two equations obtained by
cyclically permuting $\lambda, \mu, \nu$; e.g. $\dim \cU_{\mu_j}
\le \dim \cU_{\lambda_j} \dim \cU_{\nu_j}.$ The triangle
inequality then follows by applying the reasoning in the proof of
the preceding proposition.
\end{proof}

Note that our proof of the triangle inequality is very different
in spirit from the conventional one that applies subadditivity to
the purification of the state.

Finally, we show that the non-vanishing of a Kronecker coefficient
implies a relationship between entropies.

\begin{proposition}
\label{gsub} Let $\lambda, \mu, \nu \vdash k$. If
$g_{\lambda\mu\nu} \ne 0$, then $H(\bar \lambda) \le H(\bar \mu) +
H(\bar \nu)$, where $H(\bar \lambda)=-\sum_i \bar \lambda_i
\log(\bar \lambda_i)$ is the Shannon entropy of $\bar \lambda$.
\end{proposition}

\begin{proof} Kirillov has announced (\cite{Kir04}, Theorem 2.11) that
$g_{\lambda\mu\nu} \ne 0$ implies $g_{N\lambda \, N\mu \, N\nu}
\ne 0$, for any integer $N$, where $N\lambda$ means the partition
with lengths $N\lambda_i$. Since $\log(\dim \cU_{N\lambda})$ tends
to $NkH(\bar \lambda)$ for large $N$, inequality (\ref{space2})
implies the result we seek.
\end{proof}

\subsection*{V. CONCLUSIONS}

The ideas we introduce here build on concepts that were much in
vogue in the particle physics of the 1960's. The $SU(m) \times
SU(n)$ content of a representation of $SU(mn)$ expresses the types
of symmetry possible in nuclei and elementary particles, e.g. in the
multiplet theory of Wigner and in the eight-fold way of Ne'eman and
Gell-Mann \cite{Wigner36, Neeman61, Gell-Mann61, Gell-Mann62}. The
connection that we derive between these concepts and the spectra of
quantum states is novel and leads to surprisingly simple proofs of
subadditivity of the von Neumann entropy and the triangle
inequality. There are many ways to generalise these ideas, and
exploration of them is likely to be fruitful.

\subsection*{ACKNOWLEDGEMENTS}

We thank Professor Graeme Segal for inspiring conversations and Drs.
Andreas Winter and Aram Harrow for helpful comments on the
manuscript. This work was supported in part by a grant from the
Cambridge-MIT Institute, A$^*$Star Grant No.\ 012-104-0040 and the
EU under project RESQ (IST-2001-37559). MC acknowledges the support
of a DAAD Doktorandenstipendium and the U.K.~Engineering and
Physical Sciences Research Council.

\subsection*{NOTE}

After this work was completed (quant-ph/0409016) Klyachko announced
some very interesting results (quant-ph/0409113). These included a
theorem closely related to our theorem 2, and also a converse that
states that, if $g_{m \lambda, m \mu, m \nu} \neq 0$ for some positive
integer $m$, then there is a density operator with the triple of
spectra $\lambda, \mu, \nu$.

\bibliographystyle{plain}

\end{document}